\documentclass[reprint,aps,prl,letterpaper,superscriptaddress,longbibliography]{revtex4-1}
\usepackage{amssymb}
\usepackage{amsmath}
\usepackage{graphicx}
\usepackage{footnote}
\usepackage[table]{xcolor}
\usepackage[T1]{fontenc}
\usepackage[
pdfauthor={Paul D. Nation, Junho Suh, Miles P. Blencowe},
pdftitle={Ultra-Strong Optomechanics Incorporating the Dynamical Casimir Effect},
bookmarks=true,colorlinks=true,linkcolor=blue,urlcolor=blue,citecolor=blue]{hyperref}

\usepackage{tabularx,booktabs}
\newcommand{\ra}[1]{\renewcommand{\arraystretch}{#1}}
\newcommand{\dashrule}[1][black]{%
  \color{#1}\rule[\dimexpr.5ex-.2pt]{4pt}{.4pt}\xleaders\hbox{\rule{4pt}{0pt}\rule[\dimexpr.5ex-.2pt]{4pt}{.4pt}}\hfill\kern0pt%
}
\newcommand{\rulecolor}[1]{%
  \def\CT@arc@{\color{#1}}%
}

\begin{document}

\title{Ultra-Strong Optomechanics Incorporating the Dynamical Casimir Effect}

\author{P. D. Nation}
\email[E-mail: ]{nonhermitian@gmail.com}
\affiliation{Department of Physics, Korea University, Seoul 136-713, South Korea}
\altaffiliation[Present Address: ]{Northrop Grumman Corporation Electronic Systems, Aurora CO 80017 USA.}
\author{J. Suh}
\affiliation{Center for Quantum Measurement Science, Korea Research Institute of Standards and Science, Daejeon 305-340, South Korea}
\author{M. P. Blencowe}
\affiliation{Department of Physics and Astronomy, Dartmouth College, Hanover, NH 03755 USA}
\begin{abstract}
We propose a superconducting circuit comprising a dc-SQUID with mechanically compliant arm embedded in a coplanar microwave cavity that realizes an optomechanical system with a degenerate or non-degenerate parametric interaction generated via the dynamical Casimir effect.  For experimentally feasible parameters, this setup is capable of reaching the single-photon, ultra-strong coupling regime, while simultaneously possessing a parametric coupling strength approaching the renormalized cavity frequency.  This opens up the possibility of observing the interplay between these two fundamental nonlinearities at the single-photon level.
\end{abstract}
\date{\today}

\maketitle

\textit{Introduction}-- 
Recently, the field of optomechanics has undergone a period of rapid advancement, having achieved ground state cooling of a mechanical resonator \cite{chan:2011,teufel:2011b}, coherent state transfer between cavity and mechanical oscillators \cite{fiore:2011,palomaki:2013}, optomechanically induced transparency \cite{weis:2010,safavi-naeini:2011}, the generation of squeezed light \cite{safavi-naeini:2013, purdy:2013b}, position measurement precision close to the standard quantum limit \cite{anetsberger:2009,teufel:2009,purdy:2013}, and photon back-action evasion \cite{suh:2014}.  However, in all of these experiments a large optomechanical coupling is achieved by the introduction of a strong classical drive that effectively linearizes the optomechanical coupling \cite{aspelmeyer:2014}.  In contrast, taking advantage of the intrinsic optomechanical nonlinearity requires a single-photon optomechanical coupling strength $g_{0}$ that is larger than, or on the order of, the optical cavity decay rate $\kappa$.  If in addition $g_{0}$ is an appreciable fraction of the mechanical resonance frequency $\omega_{m}$, such that the combined nonlinearity parameter $g^{2}_{0}/\kappa\omega_{m}\gtrsim 1$, then the system is in the single-photon ultra-strong coupling regime characterized by cavity photon blockade \cite{nunnenkamp:2011,rabl:2011}.   This regime holds promise for achieving the long-standing goal of observing a macroscopic mechanical oscillator in a non-classical state of motion; mechanical steady-states in this regime can exhibit sub-Poissonian phonon statistics and negative Wigner functions \cite{qian:2012,nation:2013,lorch:2014}.  These effects can persist even in the presence of thermal fluctuations at experimentally accessible temperatures \cite{nation:2015}.  Recent circuit-QED realizations of the optomechanical interaction based on capacitive \cite{rimberg:2014,heikkila:2014,pirkkalainen:2015} or magnetic coupling \cite{via:2015} suggest that it may be possible to reach this ultra-strong coupling limit in the near future.

In a parallel set of investigations, the interplay between the radiation pressure nonlinearity and that of a degenerate parametric amplifier (DPA) has been of interest, where the inherent squeezing of the cavity mode gives rise to a change in cavity photon spectrum to which the optomechanical interaction is sensitive. This squeezing has be shown to aid in reaching the strong-coupling regime by enhancing the effect of a single-photon in the optical cavity \cite{lu:2015}, and can assist in the cooling \cite{huang:2009} and displacement sensitivity \cite{peano:2015} of the mechanical resonator, as well as in the generation of tripartite entangled states \cite{xuereb:2012}.  As in other studies, these results focus only on the weak coupling regime, and explore the resulting linearized optomechanical Hamiltonian.  To date, this approximation has not been a limitation, as the only optomechanical devices with both intrinsic nonlinear susceptibility and radiation pressure coupling are whispering gallery resonators (WGR) \cite{furst:2011,xuereb:2012,fortsch:2013}, where the optomechanical interaction is weak ($g_{0}/\kappa\sim 10^{-4}$); these systems operate well outside the single-photon coupling limit.  Note that it is also possible to generate an effective DPA term from the linearized optomechanical Hamiltonian in terms of the normal (polariton) modes of the system when using a drive detuned to the red-sideband \cite{borkje:2013,lemonde:2013,liu:2013}.  However, this transformation necessarily removes the optomechanical interaction.  Finding a system that is capable of reaching the single-photon, ultra-strong coupling regime while simultaneously possessing a degenerate, or non-degenerate, parametric interaction would allow for investigating the combined effect of these nonlinearities at the single-photon level.

Motivated by the desire to realize experimentally feasible systems permitting both single-photon ultra-strong optomechanical coupling and DPA interactions, in this Letter we describe an optomechanical scheme involving a dc-SQUID with mechanically compliant arm embedded in a coplanar microwave cavity that can in principle realize the single-photon ultra-strong coupling limit, while the cavity mode is simultaneously coupled to a DPA.  This scheme was originally examined in the context of optomechanical displacement detection \cite{blencowe:2007,nation:2008}, and in related subsequent experiments (but without the cavity mode) \cite{etaki:2008,poot:2010}.  Here, we take advantage of the recent demonstration of the dynamical Casimir effect (DCE) in related circuit devices using flux-modulated SQUIDs \cite{johansson:2009,wilson:2011,lahteenmaki:2013}, and the well-known connection between the DCE in a single-mode high-$Q$ cavity and a DPA \cite{nation:2012}.  In modulating an external flux applied to the SQUID loop around a fixed dc-bias, the sinusoidal variation of the SQUID effective inductance in turn modulates the frequency of the cavity mode non-adiabatically, leading to photon production from the quantum vacuum that can be viewed as a $\chi^{(2)}$ nonlinear susceptibility. An effective nonlinear medium is generated by amplifying quantum vacuum fluctuations in the cavity mode through modulating its electrical length.  Provided that the modulation frequency is much higher than the mechanical frequency, the resulting separation of timescales allows the optomechanical and parametric interactions to decouple, and an analysis of experimentally feasible parameters suggests that this system is in the single-photon ultra-strong optomechanical coupling regime $g^{2}_{0}/\kappa \omega_{m}\sim 1$, with an intrinsic DPA coupling strength that is on the order of the renormalized cavity frequency ($\sim\rm{GHz}$). 

\textit{Hamiltonian}--
\begin{figure}[b]
\includegraphics[width=8.6cm]{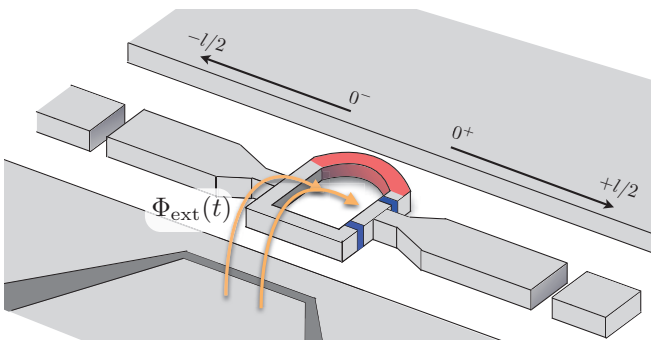}
\caption{(color online) Layout of a dc-SQUID with a mechanically compliant segment (red) embedded in a coplanar microwave cavity of length $l$, and driven by a time-dependent external flux bias $\Phi_{\rm ext}(t)$ (orange).  The SQUID, here greatly enlarged with respect to the cavity for visibility, is assumed to be a lumped element at $x=0$, with two identical Josephson junctions (blue) characterized by their critical current $I_{c}$ and capacitance $C_{J}$.}
\label{fig:fig1}
\end{figure}
The device, depicted in Fig.~\ref{fig:fig1}, consists of a coplanar microwave cavity of length $l$ and frequency $\omega_{c}$, characterized by its inductance and capacitance per unit length $L_{c}$ and $C_{c}$, respectively,  bisected by a dc-SQUID.  The SQUID comprises two Josephson junctions with critical current $I_{c}$ and capacitance $C_{J}$.  One arm of the SQUID loop is mechanically compliant, forming a doubly clamped resonator of length $l_{\rm osc}$.  We consider only the fundamental mechanical mode displacements in the plane of the SQUID loop and assume that the mechanical resonator can be modeled as a harmonic oscillator with the $y$ coordinate giving the center-of-mass displacement.  The magnetic flux threading the SQUID loop is given by $\Phi_{\rm ext}(y)=\Phi_{\rm ext}(0)+\lambda  B_{\rm ext}l_{\rm osc}y$, where $\Phi_{\rm ext}(0)\equiv \Phi_{\rm ext}$ is the flux with the mechanical oscillator fixed at $y=0$, $B_{\rm ext}$ is the local magnetic field in the vicinity of the mechanical resonator, and $\lambda$ is a dimensionless geometrical factor that accounts for the nonuniform displacement of the resonator along its extension.

To model the SQUID, we take the sum and difference of the gauge-invariant phases, $\phi_{1}$ and $\phi_{2}$,  across the Josephson junctions, $\gamma_{\pm}=(\phi_{1}\pm \phi_{2})/2$, and use the phase-field coordinate $\phi(x,t)$ for the microwave cavity.  The current and voltage along the cavity are given by the telegraph relations,
\begin{align}
I(x,t) &= -\frac{\Phi_{0}}{2\pi L_{c}}\frac{\partial \phi(x,t)}{\partial x}, \label{eq:current}\\
V(x,t) &= \frac{\Phi_{0}}{2\pi}\frac{\partial \phi(x,t)}{\partial t} \label{eq:voltage},
\end{align}
where $\Phi_{0}=h/(2e)$ is the flux quantum.  Assuming that the SQUID can be modeled as a lumped circuit element at the center $x=0$ of the cavity, the closed system equations of motion for the cavity, SQUID, and mechanical oscillator are given by \cite{blencowe:2007}
\begin{equation}\label{eq:wave}
\frac{\partial^{2} \phi}{\partial t^{2}}=\frac{1}{L_{c}C_{c}}\frac{\partial^{2}\phi}{\partial x^{2}},
\end{equation}
\begin{align}
\omega^{-2}_{J}\ddot{\gamma}_{-}&+\cos\left(\gamma_{+}\right)\sin\left(\gamma_{-}\right) \label{eq:loop_current}\\
&+ 2\beta^{-1}_{L}\left[\gamma_{-}-\pi\left(n+\frac{\Phi_{\rm ext}+\lambda B_{\rm ext}l_{\rm osc}y}{\Phi_{0}}\right)\right]=0, \nonumber \\
\omega^{-2}_{J}\ddot{\gamma}_{+}&+\sin\left(\gamma_{+}\right)\cos\left(\gamma_{-}\right)+\frac{\Phi_{0}}{4\pi L_{c}I_{\rm c}}\frac{\partial \phi(0,t)}{\partial x}=0 \label{eq:avg_current},
\end{align}
and
\begin{equation}\label{eq:newton}
m\ddot{y}+m\omega^{2}_{m}y-\frac{\Phi_{0}}{\pi L}\lambda B_{\rm ext}l_{\rm osc}\gamma_{-}=0,
\end{equation}
where $\omega_{J}=\sqrt{2\pi I_{\rm c}/(C_{J}\Phi_{0})}$ is the Josephson junction plasma frequency, $\beta_{L}\equiv 2\pi L I_{\rm c}/\Phi_{0}$ is a dimensionless parameter characterizing the SQUID self-inductance $L$, and the integer $n$ arises from the requirement that the phase around the SQUID loop must be single valued.  Equation~(\ref{eq:wave}) is the wave equation for the microwave cavity, Eq.~(\ref{eq:loop_current}) describes the current circulating through the SQUID loop, Eq.~(\ref{eq:avg_current}) gives the average current in the loop, and Eq.~(\ref{eq:newton}) is Newton's equation for the mechanical resonator driven by the Lorentz force.  The current and voltage across the SQUID must also obey the boundary conditions
\begin{equation}\label{eq:current_bc}
\frac{\partial \phi(\pm l/2,t)}{\partial x}=0, \ \ \frac{\partial \phi(0^{-},t)}{\partial x}=\frac{\partial \phi(0^{+},t)}{\partial x}
\end{equation}
and
\begin{equation}\label{eq:voltage_bc}
\dot{\gamma}_{+}-\frac{L}{4L_{c}}\frac{\partial^{2} \phi(0,t)}{\partial t \partial x}=\frac{\partial \phi(0^{-},t)}{\partial t}-\frac{\partial \phi(0^{+},t)}{\partial t},
\end{equation}
respectively.

Using Eqs.~(\ref{eq:wave})-(\ref{eq:voltage_bc}), we can derive approximate equations of motion that govern the interaction between the cavity and mechanical resonators that are determined solely by the SQUID parameters and boundary conditions.  In what follows, we assume that the SQUID plasma frequency $\omega_{J}$ satisfies $\omega_{J}\gg\omega_{c}\gg\omega_{m}$, allowing us to ignore the SQUID inertia terms in Eqs.~(\ref{eq:loop_current}) and (\ref{eq:avg_current}).  Assuming that $\beta_{L}\ll 1$ (i.e. the self-inductance of the SQUID loop is negligible),  together with dropping the SQUID inertia terms, allows us to eliminate the SQUID phase coordinates $\gamma_{\pm}$ from the equations of motion.  Furthermore, we assume that the cavity current is much less than the Josephson critical current, $|I/I_{\rm c}|\ll1$, and that the change in flux due to the small displacements of the mechanical resonator is much less than the flux quantum $|\lambda B_{\rm ext}l_{\rm osc}y/\Phi_{0}|\ll1$.  Keeping terms to first order in $y$, and to leading second order in $I$, the equation of motion (\ref{eq:newton})  for the mechanical oscillator becomes approximately
\begin{equation}\label{eq:mech_simple}
m\ddot{y}+m\omega_{m}y=\frac{I^{2}}{2}\frac{\lambda B_{\rm ext}l_{\rm osc}}{4I_{\rm c}}\sec\left(\frac{\pi\Phi_{\rm ext}}{\Phi_{0}}\right)\tan\left(\frac{\pi\Phi_{\rm ext}}{\Phi_{0}}\right),
\end{equation}
 and the voltage boundary condition (\ref{eq:voltage_bc}) can be expressed as
\begin{equation}\label{eq:new_voltage_bc}
\frac{\partial}{\partial t}\left[L\left(\Phi_{\rm ext},y\right)I\right]=\frac{\Phi_{0}}{2\pi}\left[\frac{\partial \phi(0^{-},t)}{\partial t}-\frac{\partial \phi(0^{+},t)}{\partial t}\right],
\end{equation}
with
\begin{equation}\label{eq:inductance}
L\left(\Phi_{\rm ext},y\right) = L_{J}\left(\Phi_{\rm ext}\right)\left[1+\frac{\lambda B_{\rm ext}l_{\rm osc}y}{\left(\Phi_{0}/\pi\right)}\tan\left(\frac{\pi \Phi_{\rm ext}}{\Phi_{0}}\right)\right],
\end{equation}
where we have kept only the current independent contribution to the SQUID effective inductance [the next term being of order $(I/I_{\rm c})^{2}$]
\begin{equation}
L_{J}(\Phi_{\rm ext})=\frac{\Phi_{0}}{4\pi I_{c}}\sec\left(\frac{\pi \Phi_{\rm ext}}{\Phi_{0}}\right).
\end{equation}
The validity of the approximations used in the derivation of the system equations of motion requires not approaching too close to the half-integer flux bias point:
\begin{align}
\left|\frac{I}{I_{c}}\sec\left(\frac{\pi\Phi_{\rm ext}}{\Phi_{0}}\right)\right| &\ll 1,\label{eq:current_constraint}\\
\left|\beta_{L}\sec\left(\frac{\pi\Phi_{\rm ext}}{\Phi_{0}}\right)\right| &\ll 1. \label{eq:inductance_constraint}
\end{align}
Taking the time integral of both sides of (\ref{eq:new_voltage_bc}), setting the integration constant to zero, and inserting the cavity current expression (\ref{eq:current}), Eq.~(\ref{eq:new_voltage_bc}) for the phase across the SQUID can be written as a Robin-type boundary condition:
\begin{equation}\label{eq:robin_bc}
\frac{L\left(\Phi_{\rm ext},y\right)}{L_{c}}\frac{\partial \phi(0,t)}{\partial x} = \phi(0^{+},t) - \phi(0^{-},t).
\end{equation}
In the absence of a mechanical oscillator, this boundary condition is of the same form as that used in superconducting realizations of the DCE, where the quantity $L\left(\Phi_{\rm ext},0\right)/L_{c}$ can be viewed as a flux-tunable length parameter \cite{johansson:2009, johansson:2010}.  

Let us now suppose that the external flux is weakly modulated around some fixed dc bias $\Phi_{\rm ext}=\Phi_{\rm dc}+\delta\Phi\cos(\omega_{d}t)$.  To first order in the small oscillation amplitude $\delta\Phi/\Phi_{0}$, the effective length parameter becomes
\begin{align}
\frac{L\left(\Phi_{\rm ext},y\right)}{L_{c}}\approx &\frac{L_{J}\left(\Phi_{\rm dc}\right)}{L_{c}}\left[1+\frac{\pi\delta \Phi}{\Phi_{0}}\tan\left(\frac{\pi \Phi_{\rm dc}}{\Phi_{0}}\right)\cos\left(\omega_{d}t\right)\right. \\
&+\left.\frac{\lambda B_{\rm ext}l_{\rm osc}y}{\left(\Phi_{0}/\pi\right)}\tan\left(\frac{\pi \Phi_{\rm dc}}{\Phi_{0}}\right) \right] \nonumber
\end{align}
Restricting ourselves to a single cavity mode (the extension to multiple cavity modes is straightforward), and temporarily fixing the position of the mechanical oscillator at $y=0$, then the cavity phase field that satisfies the current boundary conditions (\ref{eq:current_bc}) is
\begin{equation}\label{eq:phase}
\phi(x,t) = 
  \begin{cases}
   -\phi(t)\cos\left[k_{0}(t)\left(x+l/2\right)\right] & x<0, \\
   +\phi(t)\cos\left[k_{0}(t)\left(x-l/2\right)\right] &  x > 0,
  \end{cases}
\end{equation}
with the wavenumber $k_{0}(t)$ determined by the flux boundary condition (\ref{eq:robin_bc}):
\begin{equation}\label{eq:wavenumber}
\frac{k_{0}l}{2}\tan\left(\frac{k_{0}l}{2}\right)=\frac{L_{c}l}{L_{J}(\Phi_{\rm dc})}\left[1-\frac{\pi \delta\Phi}{\Phi_{0}}\tan\left(\frac{\pi \Phi_{\rm dc}}{\Phi_{0}}\right)\cos\left(\omega_{d} t\right)\right]
\end{equation}
For $L_{c}l/L_{J}(\Phi_{\rm dc}) \lesssim 1/2$, the left hand side of Eq.~(\ref{eq:wavenumber}) is well approximated by $(k_{0}l/2)^{2}$ that, together with $\omega^{2}_{c}=k^{2}_{0}/(L_{c}C_{c})$ from Eq.~(\ref{eq:wave}), allows the cavity mode frequency to be expressed as 
\begin{equation}\label{eq:modfreq}
\omega^{2}_{c}(t) = \left(\omega^{\rm dc}_{c}\right)^{2}\left[1-\frac{\pi \delta\Phi}{\Phi_{0}}\tan\left(\frac{\pi \Phi_{\rm dc}}{\Phi_{0}}\right)\cos\left(\omega_{d} t\right)\right],
\end{equation}
where $\left(\omega^{\rm dc}_{c}\right)^{2}=4/[C_{c}l L_{J}(\Phi_{\rm dc})]$. This expression is similar to that of Ref.~\cite{lahteenmaki:2013}, where this flux-dependent frequency modulation was used in generating radiation via the DCE in a dc-SQUID array.  The dependence of $\omega^{\rm dc}_{c}$ on the SQUID inductance (\ref{eq:inductance}) evaluated at the dc-flux bias gives a cavity frequency renormalization $\propto \cos\left(\pi\Phi_{\rm dc}/\Phi_{0}\right)$.  The plasma frequency $\omega_{J}$ is also renormalized by an amount proportional to $\sqrt{\cos\left(\pi\Phi_{\rm dc}/\Phi_{0}\right)}$, and therefore the plasma frequency remains well above $\omega^{\rm dc}_{c}$ for all flux biases compatible with Eqs.~(\ref{eq:current_constraint}) and (\ref{eq:inductance_constraint}).

Substitution of the phase field (\ref{eq:phase}) into the oscillator equation of motion (\ref{eq:newton}) gives the cavity force acting on the resonator when $y=0$.  Under our assumptions of small and slow mechanical displacements, the Lorentz force in (\ref{eq:mech_simple}) is unchanged to good approximation and the equation of motion for the mechanical oscillator becomes
\begin{align} \label{eq:mech_force}
m\ddot{y}+m\omega^{2}_{m}y=&\frac{1}{4}\left(\frac{\Phi_{0}}{2\pi}\right)^{2}C_{c}l\sin^{2}\left(\frac{k_{0}l}{2}
\right) \\
&\times \frac{\lambda B_{\rm ext}l_{\rm osc}}{\left(\Phi_{0}/2\pi\right)}\frac{L_{J}\left(\Phi_{\rm dc}\right)}{L_{c}l} \nonumber \\
&\times \tan\left(\frac{\pi \Phi_{\rm dc}}{\Phi_{0}}\right)\left(\omega^{\rm dc}_{c}\right)^{2}\phi^{2}(t), \nonumber
\end{align}
where we have assumed $\omega_{d}\gg \omega_{m}$ such that terms proportional to $\cos(\omega_{d}t)$ average to zero over a single mechanical oscillation; modulation of the applied flux does not significantly affect the coupling between the cavity and mechanical resonator.  Equation~(\ref{eq:mech_force}) allows us to determine expressions for both the mechanical portion of the system Lagrangian and that of the interaction with the cavity.  The remaining cavity terms in the Lagrangian follow from the wave equation (\ref{eq:wave}), and the total Lagrangian for the system can be written as
\begin{align}\label{eq:lagrangian}
\mathcal{L}\left(\phi,y,\dot{\phi},\dot{y}\right)&=\frac{1}{2}m\dot{y}^{2}-\frac{1}{2}m\omega^{2}_{m}y^{2}\\
&+\frac{1}{2}m_{\phi}\dot{\phi}^{2}-\frac{1}{2}m_{\phi}\omega^{2}_{c}(t)\phi^{2} \nonumber\\
&+\frac{1}{2}\frac{\lambda B_{\rm ext}l_{\rm osc}y}{(\Phi_{0}/2\pi)}\frac{L_{J}\left(\Phi_{\rm dc}\right)}{L_{c}l} \nonumber\\
&\times\tan\left(\frac{\pi\Phi_{\rm dc}}{\Phi_{0}}\right)m_{\phi}(\omega^{\rm dc}_{c})^{2}\phi^{2}, \nonumber
\end{align}
where the effective phase mass is defined to be
\begin{equation*}
m_{\phi}\equiv\frac{1}{2}C_{c}l\left(\frac{\Phi_{0}}{2\pi}\right)^{2}\sin^{2}\left(\frac{k_{0}l}{2}\right).
\end{equation*}

Defining the lowering and raising operators for the cavity mode $(\hat{a}, \hat{a}^{+})$ with respect to $\omega^{\rm dc}_{c}$, as well as those for the mechanical resonator $(\hat{b},\hat{b}^{+})$ satisfying the usual bosonic commutation relations, the system Hamiltonian takes the form
\begin{align}\label{eq:hf1}
\hat{H}&=\hbar\omega^{\rm dc}_{c}\hat{a}^{+}\hat{a}+\hbar\omega_{m}\hat{b}^{+}\hat{b}\\
&-\frac{\hbar\alpha}{2}\left(e^{+i\omega_{d}t}+e^{-i\omega_{d}t}\right)\left(\hat{a}+\hat{a}^{+}\right)^{2}\nonumber\\
&-\frac{\hbar g_{0}}{2}\left(\hat{a}+\hat{a}^{+}\right)^{2}\left(\hat{b}+\hat{b}^{+}\right)\nonumber\\
&+\hbar E \left(\hat{a} e^{+i\omega_{p}t}+\hat{a}^{+}e^{-i\omega_{p}t}\right), \nonumber
\end{align}
where, for completeness, we have included a term corresponding to the pumping of the cavity by a classical microwave field at frequency $\omega_{p}$ with amplitude $E$.  Here, the coupling strengths are expressed as
\begin{equation}\label{eq:alpha}
\alpha \equiv \frac{\omega^{\rm dc}_{c}}{4}\frac{\pi\delta\Phi}{\Phi_{0}}\tan\left(\frac{\pi \Phi_{\rm dc}}{\Phi_{0}}\right),
\end{equation}
and
\begin{equation}\label{eq:opto_coupling}
g_{0} \equiv \omega^{\rm dc}_{c}\frac{\lambda B_{\rm ext}l_{\rm osc}y_{zp}}{(\Phi_{0}/\pi)}\frac{L_{J}\left(\Phi_{\rm dc}\right)}{L_{c}l}\tan\left(\frac{\pi\Phi_{\rm dc}}{\Phi_{0}}\right),
\end{equation}
where $y_{\rm zp}=\sqrt{\hbar/(2m\omega_{m})}$ is the zero-point displacement of the mechanical resonator.  The coupling $\alpha$ is the product of both the flux-modulation amplitude [e.g. see Eq.~(\ref{eq:modfreq})] and the intrinsic parametric coupling $\chi$.  Factoring out the former, allows us to define the effective parametric coupling strength resulting from the DCE to be $\chi \equiv \omega^{\rm dc}_{c}/4$.

Moving to a frame rotating at $\omega_{d}/2$ and dropping off-resonant terms we obtain
\begin{align}\label{eq:hf2}
\hat{H} &= -\hbar\Delta\hat{a}^{+}\hat{a}+ \hbar\omega_{m}\hat{b}^{+}\hat{b}-\frac{\hbar\alpha}{2}\left[\hat{a}^{2}+\left(\hat{a}^{+}\right)^{2}\right]\\
&-\hbar g_{0}\hat{a}^{+}\hat{a}\left(\hat{b}+\hat{b}^{+}\right)+\hbar E\left(\hat{a}e^{-i\delta t}+\hat{a}^{+}e^{+i\delta t}\right),\nonumber
\end{align}
with detunings $\Delta = \omega_{d}/2-\omega^{\rm dc}_{c}$ and $\delta=\omega_{d}/2-\omega_{p}$. Equation~(\ref{eq:hf2}) describes the radiation-pressure interaction between mechanical and microwave oscillators driven by an effective DPA term arising from the DCE, as well as by a classical linear pumping term.  In addition to providing a resource for cavity mode squeezing, the DPA can also be utilized to control the mechanical resonator.  By detuning the driving frequency by an amount proportional to half the mechanical resonance frequency, $\Delta = \pm \omega_{m}/2$, it becomes energetically favorable for a pair of photons generated by the DCE to enter the cavity by emitting (absorbing) a phonon into (from) the resonator, thus heating (cooling) the mechanical mode.  Similarly, detuning by $\Delta = \pm\omega_{m}$ makes this emission or absorption a two phonon process.

Although we have restricted ourselves to a single cavity mode, the extension to multiple modes is straightforward.  In the case of two cavity modes,  modulating the cavity resonance frequency at the sum of the two dc-biased frequencies gives rise to a resonant non-degenerate parametric coupling $(\hat{a}\hat{c}+\hat{a}^{+}\hat{c}^{+})$, where the operators $(\hat{c},\hat{c}^{+})$ correspond to a cavity mode with non-vanishing current at the location of the SQUID, i.e. the cavity mode phase must satisfy $\partial\phi/\partial x†\neq 0$ at $x=0$.  In addition, driving at the difference of the mode frequencies results in a beam-splitter interaction $(\hat{a}\hat{c}^{+}+\hat{a}^{+}\hat{c})$, which has also been investigated for membrane in the middle configurations \cite{ludwig:2012,stannigel:2012}.  With the mechanical resonator coupled to both modes via radiation pressure coupling, possibly in the single-photon strong-coupling limit, such a configuration provides a rich testbed for investigating tripartite entanglement in strongly nonlinear systems \cite{tian:2013,wang:2015}.

\textit{Ultra-strong coupling regime}--
Given that the optomechanical coupling is dependent on the external magnetic field, it is advantageous to consider a device fabricated using niobium, with critical field $B_{c}\sim 198~\rm{mT}$, rather than aluminum ($B_{c}\sim 10~\rm{mT}$).  A set of experimentally feasible parameters compatible with this choice is: $\Phi_{\rm dc}=0.35\Phi_{0}$, $\omega^{\rm dc}_{c}(0)=2\pi\times 10~\rm GHz$, $\omega_{m}=2\pi\times 10~\rm MHz$, $m=10^{-16}~\rm kg$, $l_{\rm osc}=10~\mu\rm m$, $\lambda=2/\pi$, $B_{\rm ext}=40~\rm mT$, $I_{c}=100~\rm nA$ and $L_{c}l=1~\rm nH$.  At this dc-bias, the SQUID renormalizes the cavity frequency to nearly half of its original value $\omega^{\rm dc}_{c}\simeq 2\pi\times 4.5~\rm GHz$.  In addition, we assume quality factors for the dc-biased cavity and mechanical resonator to be $Q^{\rm{dc}}_{c}=5 \times 10^{4}$ and $Q_{m}=10^{4}$, respectively.  The former value has already been achieved in flux-tunable Al resonators \cite{kubo:2010}, although the addition of a flux bias degrades the quality factor in these oscillators.  In contrast, the quality factor for Nb cavities is marginally affected by an external bias provided that the renormalized frequency is large enough for thermal effects to be ignored \cite{palacios-laloy:2008}.  For the frequencies considered here, the thermal occupation of the cavity mode is $\sim 10^{-10}$ at $10~\rm{mK}$, and thus we expect a negligible change in the quality factor.   Substitution of these parameters into Eq.~(\ref{eq:opto_coupling}) gives $g_{0}/\kappa\approx 13$ and $g^{2}_{0}/\kappa \omega_{m}\approx 1.3$, indicating that this setup lies within the single-photon ultra-strong coupling regime. In addition, the intrinsic parametric coupling strength at this dc-bias is $\chi = 2\pi \times 1~\rm GHz$.  The total parametric coupling strength $\alpha$ (\ref{eq:alpha}) will be much lower due to the weak modulation amplitude $\delta\Phi/\Phi_{0}\ll 1$ and the stability requirements \cite{huang:2009, jimenez:2014} of the system.  A comparison of these figures of merit with recent experimental and theoretical optomechanical systems incorporating macroscopic mechanical resonators is presented in Table~\ref{table}.

\begin{table}\centering
\ra{1.1}
\begin{tabularx}{\columnwidth}{
>{\raggedright\arraybackslash\advance\hsize8em}X
>{\raggedright\arraybackslash\advance\hsize0.7em }X
>{\raggedright\arraybackslash\advance\hsize0.7em}X
>{\raggedright\arraybackslash\advance\hsize0.7em }X
>{\raggedright\arraybackslash\advance\hsize-0em }X
}
\toprule
\large System & $\dfrac{g_{0}}{\kappa}$ & $\dfrac{g_{0}}{\omega_{m}}$ & $\dfrac{g^{2}_{0}}{\kappa\omega_{m}}$ & $\dfrac{\chi}{2\pi}$ \\
\midrule
Microwave LC-drum \cite{palomaki:2013} & $6\cdot 10^{-4}$ & $2\cdot 10^{-5}$ & $10^{-8}$ & n/a \\
\addlinespace[0.25em]
Si zipper cavity \cite{safavi-naeini:2013} & $2\cdot 10^{-4}$ & $3\cdot 10^{-2}$ & $6\cdot 10^{-6}$ &  n/a \\
\addlinespace[0.25em]
Circuit-QED qubit \cite{pirkkalainen:2015} & $4\cdot 10^{-2}$ & $2\cdot 10^{-2}$ & $8\cdot 10^{-4}$ &  n/a \\
\multicolumn{5}{@{}c@{}}{\makebox[\linewidth]{\dashrule[black!100]}} \\[-\jot]
cCPT - resonator \cite{rimberg:2014} & $10$ & $1$ & $10$ &  n/a\\
\addlinespace[0.25em]
Stripline - cantilever \cite{via:2015} & $20$ \footnote{Ref.~\cite{via:2015} assumes a cavity quality factor of $10^{6}$ taken from $3D$ superconducting cavity geometries rather than the stripline configuration considered in that work.  As such, the realizable values for $g_{0}$ are likely to be lower.} & -- & -- &  n/a \\
\addlinespace[0.25em]
$\rm{Si}_{3}\rm{N}_{4}$ WGR \cite{xuereb:2012}& $2\cdot 10^{-4}$ & $2\cdot 10^{-5}$ & $4\cdot 10^{-9}$ & $110~\rm{Hz}$ \\
\addlinespace[0.25em]
\textbf{SQUID - resonator} & $\bf{13}$  & $\bf{0.1}$ & $\bf{1.3}$ & $\bf{1~\rm {GHz}}$ \\
\bottomrule
\end{tabularx}
\caption{Ratio of the optomechanical coupling strength to that of the cavity decay rate, $g_{0}/\kappa$ and mechanical frequency $g_{0}/\omega_{m}$, combined quantum nonlinearity parameter $g^{2}_{0}/\kappa\omega_{m}$, and intrinsic parametric coupling strength $\chi$ for a selection of recent experimental (above dashed-line) and theoretical (below dashed-line) optomechanical systems incorporating macroscopic mechanical resonators.}
\label{table}
\end{table}

Although we have tacitly assumed that our optomechanical system involves only a single mode of the mechanical oscillator, in practice all odd harmonics that are symmetric about the midpoint of the resonator will couple to the SQUID through the modulation of the external flux.  Our single-mode approximation rests on the assumption that higher-order modes are spectrally well separated from the fundamental. However, given the large single-photon coupling strength in this device, it is of interest to ask whether additional higher-modes are also strongly coupled to the cavity.  The tripartite interaction between a cavity and two mechanical modes has already been explored for generating entanglement \cite{huang:2009b,massel:2012}, two-mode squeezing \cite{woolley:2014}, and EPR states \cite{zhang:2003}, while as an ensemble of mechanical modes has been proposed for quantum information processing \cite{schmidt:2012}.  Assuming that the higher modes are harmonics of the fundamental, the radiation pressure coupling (\ref{eq:opto_coupling}) is reduced by a factor of $1/n^{3/2}$, where $n$ labels the harmonic, arising from a reduction in both the geometrical factor accounting for the net change in SQUID loop area and a decrease in the zero-point amplitude at higher frequencies (assuming the mass of each mode is approximately the same).  For the parameters detailed here, this estimation suggests that up to the first six odd-harmonics of the resonator can be simultaneously strongly (but not ultra-strongly) coupled to the microwave cavity.

\textit{Conclusion}-- 
We have shown that a dc-SQUID with mechanically compliant segment embedded in a coplanar microwave cavity modulated by a time-dependent external flux gives rise to an optomechanical interaction where the cavity mode is coupled to a degenerate or non-degenerate parametric amplifier.  Using experimentally feasible device parameters  indicates that this setup is capable of operating in the single-photon ultra-strong optomechanical coupling regime, while simultaneously possessing a degenerate parametric coupling strength on the order of the dc-bias renormalized cavity frequency.  Additional higher-modes of the mechanical resonator can be strongly-coupled simultaneously in this setup.  In contrast to previous proposals for strongly-coupled optomechanics, where the frequency of the mechanical resonator is intrinsically low \cite{via:2015}, or is renormalized downward \cite{rimberg:2014,heikkila:2014} such that thermal effects conspire to degrade the quantum signatures in these setups, the present scheme lies within the parameter space where quantum signatures such as negative Wigner functions are predicted to persist at standard dilution refrigerator temperatures \cite{nation:2015}, and thus is well-suited for investigating macroscopic quantum states of the mechanical resonator.   Compared to current WGR systems, the device presented here gives a four and nine order of magnitude increase in $g_{0}/\kappa$ and $g^{2}_{0}/\kappa\omega_{m}$, respectively.  Combined with the strong intrinsic parametric coupling $\chi$, this setup opens the door to exploring the expected rich physics arising from the interaction between these two principle nonlinearities, in a single or multi-mode configuration, at the individual photon level.

The authors thank J. R. Johansson and N. Lambert for helpful discussions.  P.D.N. was supported by the MSIP (Ministry of Science, ICT and Future Planning) Korea under the ITRC (Information Technology Research Center) support program (IITP-2015-R0992-15-1017), supervised by the IITP (Institute for Information \& Communications Technology Promotion), and startup funding from Korea University. J.S. was supported by the KRISS project `Convergent Science and Technology for Measurements at the Nanoscale'. M.P.B. was supported by the NSF Grant No. DMR-1104790.

\bibliography{refs}
\end{document}